\documentclass[pra,twocolumn,showpacs]{revtex4-1}

\usepackage{amsmath}
\usepackage{graphicx}

\newcommand\myup{\mathord{\uparrow}}
\newcommand\mydown{\mathord{\downarrow}}

\begin{document}
\title{Momentum-resolved spectroscopy of a Fermi liquid}
\author{Elmer V. H. Doggen}
\author{Jami J. Kinnunen}
\email[Corresponding author: ]{jami.kinnunen@aalto.fi}
\affiliation{COMP Centre of Excellence and Department of Applied Physics, Aalto University, FI-00076 Aalto, Finland}

{\abstract We consider a recent momentum-resolved radio-frequency spectroscopy experiment, in which Fermi liquid properties of a strongly interacting atomic Fermi gas were studied.
Here we show that by extending the Brueckner-Goldstone model, we can formulate a theory that goes beyond basic mean-field theories and that can be used for studying spectroscopies of dilute atomic gases in the strongly interacting regime.
The model hosts well-defined quasiparticles and works across a wide range of temperatures and interaction strengths.
The theory provides excellent qualitative agreement with the experiment.
Comparing the predictions of the present theory with the mean-field Bardeen-Cooper-Schrieffer theory yields insights into the role of pair correlations, Tan's contact, and the Hartree mean-field energy shift.}

\maketitle

Strongly interacting fermionic systems are ubiquitous in nature; they are found from solid state systems \cite{Mirzaei2013a} and fermionic superfluids to neutron stars and nuclear matter.
In the context of ultracold atoms, the transition from weak to strong interactions is described by the crossover from Bardeen-Cooper-Schrieffer (BCS) theory to a Bose-Einstein condensate (BEC) of pairs of fermions \cite{Randeria2014a}.
In between these two regimes of weakly interacting particles, the system is in the \emph{unitary} regime \cite{Thomas2012a}, where the absence of a small parameter makes standard perturbation theory inadequate.
These systems are therefore more difficult to describe theoretically.
In the highly controllable environment of ultracold atoms, one can tune the interactions using Feshbach resonances \cite{Kokkelmans2014a}, making the BCS-BEC crossover accessible in the experiment.

On the BCS side of the crossover, the system is found in a superfluid state below a certain critical temperature $T_\mathrm{c}$, where BCS theory is applicable.
In this state, fermions form so-called Cooper pairs in momentum space.
Above $T_\mathrm{c}$, in the \emph{normal state}, the pairs are not formed, and the system is found to be described well as a Fermi liquid~\cite{Nascimbene2011a}.
In a Fermi liquid, the system behaves similar to a non-interacting gas of fermions, with well-defined and long-lived fermionic quasiparticles which have an effective mass.
In this phase, the momentum distribution has a ``jump'' of size $Z$ at the Fermi momentum $k_\mathrm{F}$.
The value of $Z$, the \emph{quasiparticle weight}, depends on both the sign (attractive or repulsive) and magnitude of the interactions, and its vanishing corresponds to the breakdown of the Fermi liquid description as investigated in a recent experiment at JILA~\cite{Sagi2014a}.

\begin{figure}
\centering
\includegraphics[width=0.9\columnwidth]{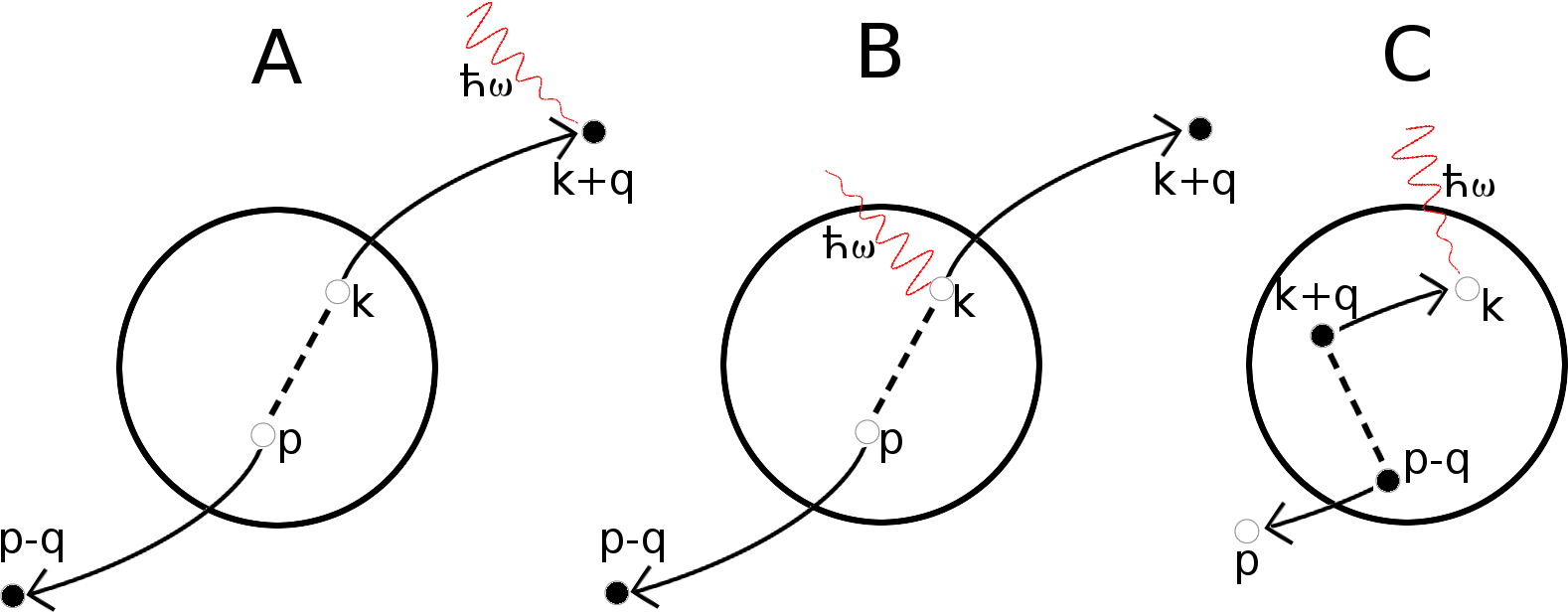}
\caption{Physical scattering processes described by the perturbative corrections to the Brueckner-Goldstone response function $S_\mathrm{k}(\delta)$. See main text. Diagram A describes a process in which two particles with momenta $+$ and $-$ are scattered to momenta $k$ and $p$ and the radio-frequency photon of energy $\hbar \omega$ flips the spin-state of the momentum $k$ atom. Diagram B describes a shadow process of diagram A, in which the atoms scatter away from states $k$ and $p$, leaving holes in place. Finally, the diagram C describes a process in which the rf-photon first excites an atom with momentum $k$, leaving thus a hole in the sea of $|\myup\rangle$-atoms. 
This is followed by the scattering of two atoms with momenta $+$ and $-$ into the hole in $k$ and some empty state $p$.}
\label{fig:diagrams}
\end{figure}

A convenient experimental technique for studying ultracold atoms is radio-frequency spectroscopy, which has been applied in many experimental as well as theoretical approaches~\cite{Chen2009a, Levin2012a}.
Radio-frequency spectroscopy can, for instance, be used to obtain the \emph{contact} \cite{Tan2008a,Tan2008b,Tan2008c,Zwerger2012a,Stewart2010a}, a quantity describing short-range correlations in the system.
By measuring momenta of the atoms transferred by the long wavelength radio-frequency field~\cite{Stewart2008a,Chen2009a,Gaebler2010a}, one can determine the single-particle spectral function of the atoms in the initial many-body state~\cite{Gaebler2010a}.
Furthermore, by selectively probing the system so that one considers only a particular ``slice'' where the density is approximately homogeneous \cite{Jin2012a}, the method allows experimental verification of theories in the unitary regime.

The theory used in this work for describing the BCS-BEC crossover is a perturbative extension of the Brueckner-Goldstone (BG) theory~\cite{FetterAndWalecka, Brueckner1955a, Goldstone1957a}, which has primarily been applied in the context of nuclear physics and liquid $^3$He~\cite{Glyde1983a}.
This theory is similar to Fermi liquid theory in the sense that it has long lived quasiparticles at the Fermi surface, and an associated jump in the momentum distribution. 
This is in contrast to BCS theory, in which the formation of pairs results in a continuous momentum distribution.
Well-formed pairs are a given in the superfluid phase of the Fermi gas, as well as in the BEC side of the BCS-BEC crossover in which two-body physics supports a (molecular) bound state.
However, bound states are not always antithesis to Fermi liquid-like behavior \cite{Engelbrecht1992a,Cazalilla2011a,Koschorreck2012a}.
The goal of the present work is to study to what extent Fermi liquid picture can be used in strongly interacting atomic gases.
In particular, we describe a situation in which pairing is \emph{not} important, and we instead focus on scattering processes between the atoms.
The breakdown of the theory can then be associated with the appearance of pairs, giving physical intuition into which processes dominate the system.
This can be seen as an approach complementary to BCS theory, which assumes pairs and breaks down when the pairs become unstable to decay, or as an alternative to many pseudogap theories~\cite{Levin2012a,Perali2002a,Hu2010a,Wlazlowski2013a} in which noncondensed pairs are formed already at temperatures above the critical superfluid temperature.

BG theory is appealing for various reasons.
It can be formulated in terms of the more well-behaved two-body scattering T-matrix, rather than the bare inter-atomic potential.
Furthermore, the theory can describe the Hartree energy shift even at unitarity where the na\"ive (mean-field) constant energy shift $\frac{4\pi\hbar^2a}{m} n$ diverges as the scattering length $a \rightarrow \infty$ (where $n$ is the atom density and $m$ is the mass of the atom). 
The model also provides, as a by-product, the full many-body scattering T-matrix, which, in turn, can be used for extending the model.
Here we will extend the BG theory perturbatively, and use it for
calculating the momentum-resolved radio-frequency response function.
The perturbative processes included in the response function are shown as schematic diagrams in Fig.~\ref{fig:diagrams}.

\section{Results}

{\bf Hartree shift and effective masses.} 
The interacting two-component Fermi gas is described by the many-body Hamiltonian
\begin{align}
 & \hat H_\mathrm{int} = -\sum_{\sigma} \int d{\bf r}\, \hat \psi_\sigma^\dagger({\bf r}) \frac{\hbar^2}{2m} \nabla^2 \hat \psi_\sigma({\bf r}) \nonumber \\
 & + \frac{1}{2}\int d{\bf r} \int d{\bf r'} \hat \psi_\uparrow^\dagger ({\bf r}) \hat\psi_\downarrow^\dagger ({\bf r'}) \hat V({\bf r-r'}) \hat \psi_\downarrow ({\bf r'}) \hat \psi_\uparrow({\bf r}),
\end{align}
where $m$ is the mass of an atom, assumed to be equal for all (pseudo)spin states, and $\hat \psi_\sigma^{(\dagger)}({\bf r})$ is a field operator, which annihilates (creates) an atom with (pseudo)spin $\sigma \in \{\uparrow,\downarrow,e\}$ at point ${\bf r}$. 
The different components, or (pseudo)spin states, correspond to different hyperfine states of the atoms.
In the presence of a magnetic field, these different internal states of the atoms are well defined with large energy gaps due to Zeeman effect.
In dilute and cold atomic gases, the hyperfine states form an excellent analogy of spin-N (for bosonic atoms) or spin-N+$\frac{1}{2}$ (for fermionic atoms) particles.
For simplicity, we will refer to these different hyperfine states as spin-states.
The atoms are assumed to be fermionic and, consequently, the field operators satisfy anticommutation relations $\left\{ \hat \psi_\sigma^\dagger ({\bf r}), \hat \psi_{\sigma'} ({\bf r'})\right\} = i \delta({\bf r-r'}) \delta_{\sigma,\sigma'}$.
The two-particle interaction potential $V({\bf r})$ is assumed to be of short range, in which case its details are irrelevant. 
However, the two-body scattering T-matrix used below corresponds to the contact interaction pseudopotential $\hat V({\bf r}) = V_0 \delta({\bf r}) \frac{d}{dr} \left( r \cdot \right)$, where $V_0 = \frac{4\pi \hbar^2 a}{m}$ and $a$ is the $s$-wave scattering length.
Notice that the model involves three different hyperfine states of the atoms $|\myup\rangle$, $|\mydown\rangle$, and $|e\rangle$: the initial state is a balanced mixture of $|\myup\rangle$ and $|\mydown\rangle$ atoms, and the radio-frequency field transfers atoms from the state $|\myup\rangle$ to the initially unoccupied non-interacting state $|e\rangle$.

The coupling with the probing radio-frequency (rf) field is described by the standard time-dependent operator in the rotating wave approximation
\begin{equation}
  \hat H_\mathrm{rf} = \Omega e^{i\delta t} \int d{\bf r} \, \hat \psi_e^\dagger({\bf r}) \hat \psi_\uparrow ({\bf r}) + \text{H.c.},
\end{equation} 
where $\Omega$ is the coupling strength and $\delta$ is the frequency detuning of the rf-photon from the hyperfine energy splitting between spin states $|e\rangle$ and $|\myup\rangle$.

The transfer rate for atoms with momentum ${\bf k}$ in hyperfine state $|\myup\rangle$ to be transferred to the $|e\rangle$-state at time $\tau$ by the rf-pulse is given by linear response theory as
\begin{equation}
  S_k(\delta) = 2 \mathrm{Im}\, \int_{-\infty}^\infty \frac{d\omega}{2\pi} G_\uparrow({\bf k},\omega) G_e({\bf k},\omega-\delta).
\label{eq:spectrum}
\end{equation}
This spectral function is normalized in such a way that integration over frequency $\delta$ yields the occupation probability $n(k)$ of the momentum state $k$ in the initial state.
Because atoms in the excited state $|e\rangle$ are noninteracting, and initially there are no atoms in the state, the corresponding Green's function has the simple form of a vacuum propagator
\begin{equation}
   G_e({\bf k},\omega) = G_0({\bf k},\omega)= \frac{\hbar}{\hbar\omega-\epsilon_k+i\eta},
\label{eq:nonintG}
\end{equation}
where $\epsilon_k = \frac{\hbar^2 k^2}{2m}$ and $\eta$ is a convergence parameter. 
What is needed now is the Green's function for atoms in the spin state $|\myup\rangle$.

The Brueckner-Goldstone theory, outlined in the Methods section, provides a good basis for formulating a theory that can incorporate many-body interactions across the BCS-BEC crossover.
The starting point is the Dyson equation, which connects the interacting Green's function and the self-energy $\Sigma_\uparrow$:
\begin{equation}
  G_\uparrow({\bf k},\omega)^{-1} = G_0({\bf k},\omega)^{-1} - \Sigma_\uparrow({\bf k},\omega).
  \label{eq:dyson}
\end{equation}
Different approximations to the self-energy then yield different many-body theories~\cite{Haussmann1993a,Haussmann1994a,Levin2012a,Perali2002a,Haussmann2007a,Watanabe2010a,Hu2010a,Gubbels2011a,Wlazlowski2013a}.
In Brueckner-Goldstone theory, one considers only self-energies on-the-energy-shell (or simply on-shell), i.e. the energy dependent part of the self-energy is neglected and evaluated at the energy equal to the interacting single-particle energy:
\begin{equation}
  \Sigma_\mathrm{BG} (k) = \Sigma_\uparrow(k,\epsilon_k + \Sigma_\mathrm{BG}(k)).
\label{eq:BGselfenergy}
\end{equation}
We will provide the perturbative extension of the Brueckner-Goldstone theory in a moment, but it is worthwhile to consider already the behavior of the Brueckner-Goldstone self-energy $\Sigma_\mathrm{BG}(k)$ itself. It allows us to calculate several experimentally relevant quantities, such as the Hartree energy shift and effective masses.

\begin{figure}
\centering
\includegraphics[width=0.9\columnwidth]{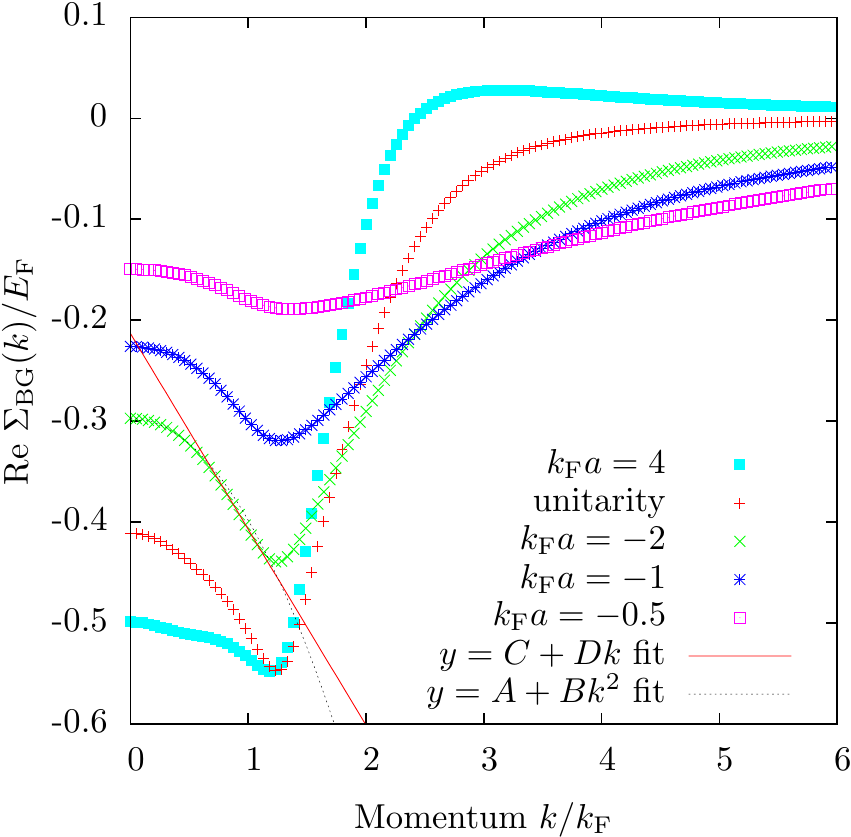}
\caption{The real part of the Brueckner-Goldstone self-energy $\Sigma_\mathrm{BG}$ as a function of momentum for various interaction strengths.  
Shown are also a $y=A+Bk^2$ fit to the $k_\mathrm{F}a=-2$ self-energy data in the range $k/k_\mathrm{F} \in [0,0.5]$ used for determining the zero momentum effective mass and a linear $y=C+Dk$ fit for the data in the range $k/k_\mathrm{F} \in [0.9,1.1]$ for obtaining the effective mass at the Fermi surface.
Here and elsewhere, unless otherwise pointed out, the temperature is $T = 0.2\,T_\mathrm{F}$ and the convergence factor $\eta = 0.05\,E_\mathrm{F}$.}
\label{fig:hartree}
\end{figure}

Fig.~\ref{fig:hartree} shows the calculated real part of the self-energy $\Sigma_\mathrm{BG}(k)$ for various interaction strengths.
The plot reveals the strong momentum dependence of the self-energy, particularly close to unitarity $k_\mathrm{F}a = \pm \infty$.
The momentum dependence is easily understood~\cite{Kinnunen2012a} when considering the two-body on-shell scattering amplitude, which for the contact interaction pseudopotential is
\begin{equation}
   f(k) = \frac{a}{1+ika}.
\end{equation}
For large relative momenta $k \gg 1/a$, the scattering amplitude is suppressed. 
Hence, high momentum atoms will interact very weakly with atoms in the Fermi sea and the self-energy is suppressed.
Deep inside the Fermi sea for $k \ll k_\mathrm{F}$, the self-energy is again suppressed. 
This is caused by the Pauli blocking of low-energy scattering channels due to the Fermi sea. 
Subsequently the self-energy has a (negative) maximum close to the Fermi surface. 
In the weakly interacting limit $|k_\mathrm{F}a| \ll 1$, the real-part of the self-energy reproduces the usual Hartree energy shift $\frac{4\pi\hbar^2 a}{m} n_\sigma$, where $n_\sigma$ is the density of atoms in one spin state. 
In this limit, the momentum dependence of the scattering amplitude is also insignificant since it will not play a role until momenta $k \gg 1/a$.

\begin{figure}
\centering
\includegraphics[width=0.9\columnwidth]{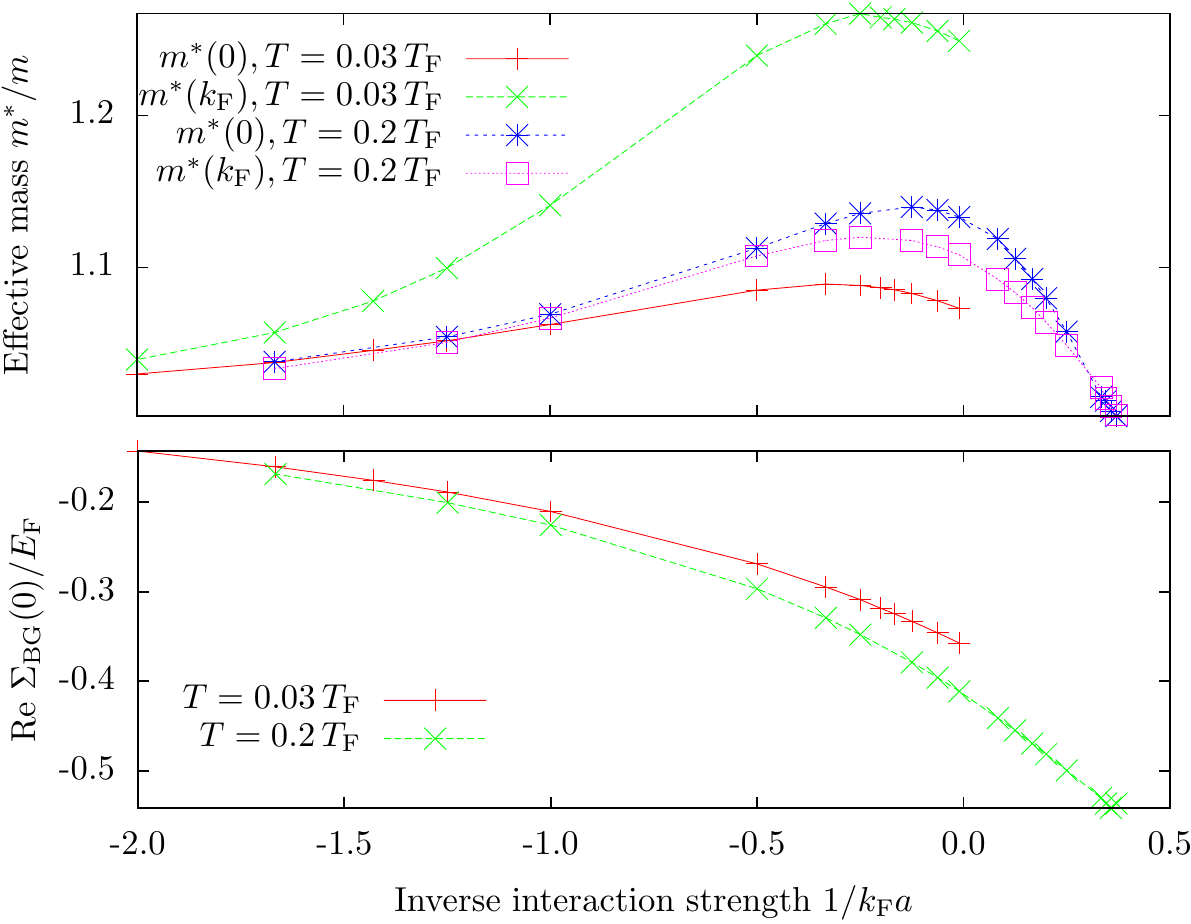}
\caption{Top: the effective mass $m^*/m$ obtained from the Brueckner-Goldstone self-energy for zero momentum atoms and atoms at the Fermi surface. 
The zero momentum effective mass is obtained using a quadratic fit to the self-energy, and the effective mass at the Fermi surface using a linear fit as exemplified in Fig.~\ref{fig:hartree}. The $T=0.2\,T_\mathrm{F}$ data for $k=0$ shows the error bars from the fitting. Bottom: the energy shift of zero-momentum atoms $\mathrm{Re}\,\Sigma_\mathrm{BG}(k=0)$ as a function of interaction strength. Notice that the data for different temperatures has not been calculated beyond the point where the perturbative extension of the Brueckner-Goldstone model starts exhibiting nonphysical artifacts in the momentum distribution, see main text. The model works better at higher temperatures.}
\label{fig:effmass}
\end{figure}

The momentum dependence of the self-energy implies that quasiparticles behave as having an effective mass $m^*$, which can differ from the bare atom mass $m$. 
The effective mass depends on momentum, and for a given momentum $k$ it can be determined by fitting quadratic dispersion to the dispersion of the quasiparticle as follows
\begin{equation}
   \varepsilon_k = \frac{\hbar^2 k^2}{2m} + \mathrm{Re}\, \Sigma_\mathrm{BG}(k) \approx_\mathrm{fit} \frac{\hbar^2 k^2}{2m^*}.
\end{equation} 
In practice, this means doing a linear, or quadratic if $k=0$, fit to the self-energy, as exemplified in Fig.~\ref{fig:hartree}.
In particular, the zero-momentum effective mass is
\begin{equation}
  \frac{m^*}{m} = \frac{\hbar^2}{m} \left.\left(\frac{\partial^2 \varepsilon_k}{\partial k^2} \right)^{-1} \right|_{k=0},
\end{equation}
and at the Fermi momentum it is
\begin{equation}
  \frac{m^*}{m} = \frac{\hbar^2 k_\mathrm{F}}{m} \left.\left(\frac{\partial \varepsilon_k}{\partial k} \right)^{-1} \right|_{k=k_\mathrm{F}}.
\end{equation}
These effective masses are shown as a function of interaction strength in Fig.~\ref{fig:effmass}.
Interestingly, the figure shows a clear maximum in the strongly interacting regime, away from unitarity.
Both of the effective masses have the same qualitative behavior, although the effect is more pronounced at the Fermi surface because interaction effects are stronger at the Fermi surface.
The decreasing effective mass when crossing the unitary limit can be understood as a crossover to a repulsive single-particle branch. 
While the ground state in the BEC side consists of molecules, with effective mass $m^* = 2m$ in the far BEC limit, the
unpaired fermions will in the same limit have effective mass of $m^* = m$, because the single-particle branch and molecular branch become separated by a large energy gap.
The present theory is unable to describe the molecular branch, but it should provide a good description of repulsively interacting unpaired fermions sufficiently far in the BEC limit.

An interesting effect is the temperature dependence of effective masses. 
The effective mass of zero momentum atoms increases with higher temperature while for atoms at the Fermi surface it decreases.
The first effect is due to the appearance of thermal hole excitations within the Fermi sea, opening up some of the low-energy scattering channels that would otherwise have been blocked. 
This increases the effective interaction strength of low momentum atoms.
In contrast, atoms at the Fermi surface have decreased scattering probability because the Cooper instability, which describes many-body enhancement of scattering processes around the Fermi surface, is weakened with the broadening of the Fermi surface.

Fig.~\ref{fig:effmass} shows also the energy shift of zero-momentum atoms, $\mathrm{Re}\, \Sigma_\mathrm{BG}(0)$. 
It shows smooth behavior near unitarity, although sufficiently deep in the BEC side the self-consistent iteration has problems finding a unique solution. 
Close to $k_\mathrm{F}a \approx 2$ the model switches to the repulsive single-particle branch, involving a big change in the self-energy.
While we consider this to be due to the limitations of the model, namely that it cannot simultaneously describe both the repulsive single-particle branch and the molecular branch, it is intriguing that the experiment~\cite{Sagi2014a} also exhibits a sudden change to the repulsive branch at a comparable interaction strength.

{\bf Momentum distribution, contact, and quasiparticle weight.} In order to analyze momentum distributions and spectral functions, the Brueckner-Goldstone theory must be extended.
Indeed, the on-shell approximation for the self-energy made in Eq.~\eqref{eq:BGselfenergy} yields no corrections to the non-interacting distributions.
However, we can use the Dyson equation~\eqref{eq:dyson} for formulating a perturbative correction to the interacting Green's function as
\begin{align}
   & G(k,\omega) \approx  G_\mathrm{BG}(k,\omega) + G_\mathrm{BG}(k,\omega) \Big[ \Sigma(k,\omega) \nonumber \\
   & - \Sigma_\mathrm{BG}(k) \Big] G_\mathrm{BG}(k,\omega) =: G_\mathrm{pert}(k,\omega). \label{eq:pertG}
\end{align}
As shown in Section~\ref{sec:methods}, using the perturbed Green's function $G_\mathrm{pert}({\bf k},\omega)$ yields the momentum distribution
\begin{eqnarray}
   n(k) &=& n_k + \int \frac{d{\bf p}d{\bf q}}{(2\pi)^6} \left|\Gamma_\mathrm{os}\right|^2 \frac{\left(1-n_k\right)\left(1-n_p\right) n_+ n_-}{\left(\varepsilon_+ + \varepsilon_- - \varepsilon_k^* - \varepsilon_p^*\right)^2} \nonumber\\
&-& \int \frac{d{\bf p}d{\bf q}}{(2\pi)^6} \left|\Gamma_\mathrm{os}\right|^2 \frac{n_k n_p \left(1-n_+\right) \left(1- n_-\right)}{\left(\varepsilon_k + \varepsilon_p - \varepsilon_+^* - \varepsilon_-^*\right)^2}.
\label{eq:momcorr}
\end{eqnarray}
where $\Gamma_\mathrm{os}$ is the on-shell scattering T-matrix and the subscripts $\pm$ correspond to momenta ${\bf (k+p)}/2 \pm {\bf q}$.
The first term is simply the unperturbed occupation probability (the Fermi-Dirac distribution at fixed temperature).
The other terms are the perturbative correction to the momentum distribution:
the second term gives the probability that a particle has scattered to an initially empty state with momentum ${\bf k}$, and
the third term is the probability that an initially occupied state is empty, due to scattering to other states.
The perturbative correction can be shown to conserve the number of particles, although it is not guaranteed to yield occupation numbers between 0 and 1 below the superfluid phase transition temperature.
This anomalous feature is not surprising, given that we are explicitly neglecting superfluid pairing \emph{a priori}.
However, the momentum distribution is well-behaved even at unitarity when the temperature is sufficiently high ($T\gtrsim0.2\,T_\mathrm{F}$).
In the weakly interacting limit, Eq.~\eqref{eq:momcorr} reproduces analytical results of Ref.~\cite{Sartor1980a}. 

\begin{figure}[!htb]
\centering
\includegraphics[width=0.9\columnwidth]{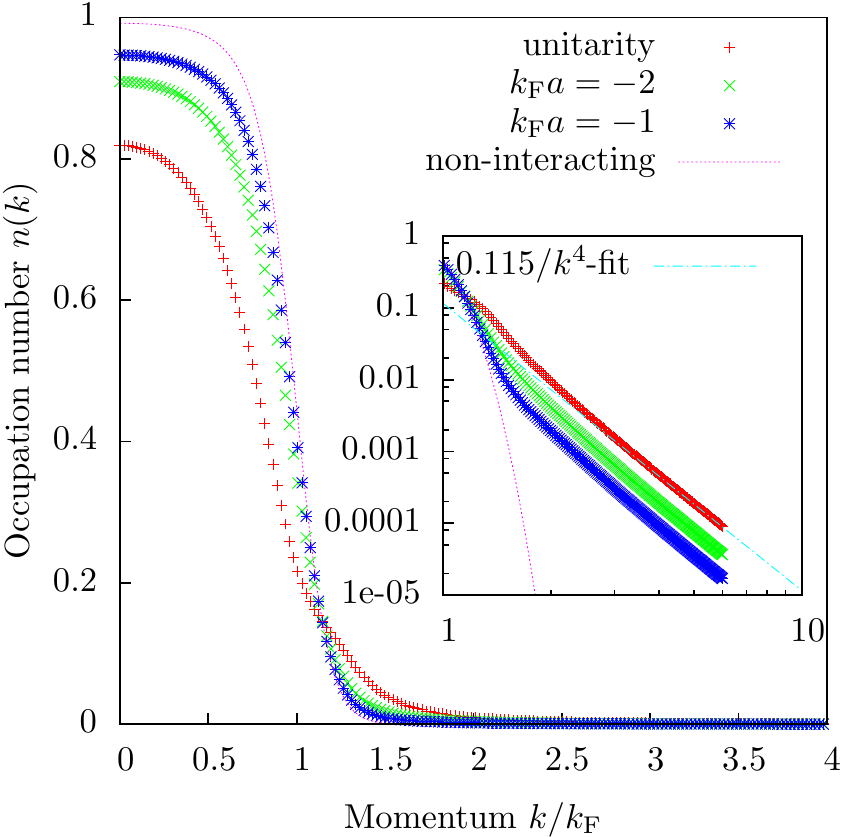}
\caption{The momentum distribution from the perturbatively extended BG theory for temperature $T=0.2\,T_\mathrm{F}$. Inset shows the same data in logarithmic scale. 
The high momentum asymptote follows $1/k^4$ scaling, which is the result of short-range interactions.}
\label{fig:momdist}
\end{figure}

\begin{figure}[!htb]
\centering
\includegraphics[width=0.9\columnwidth]{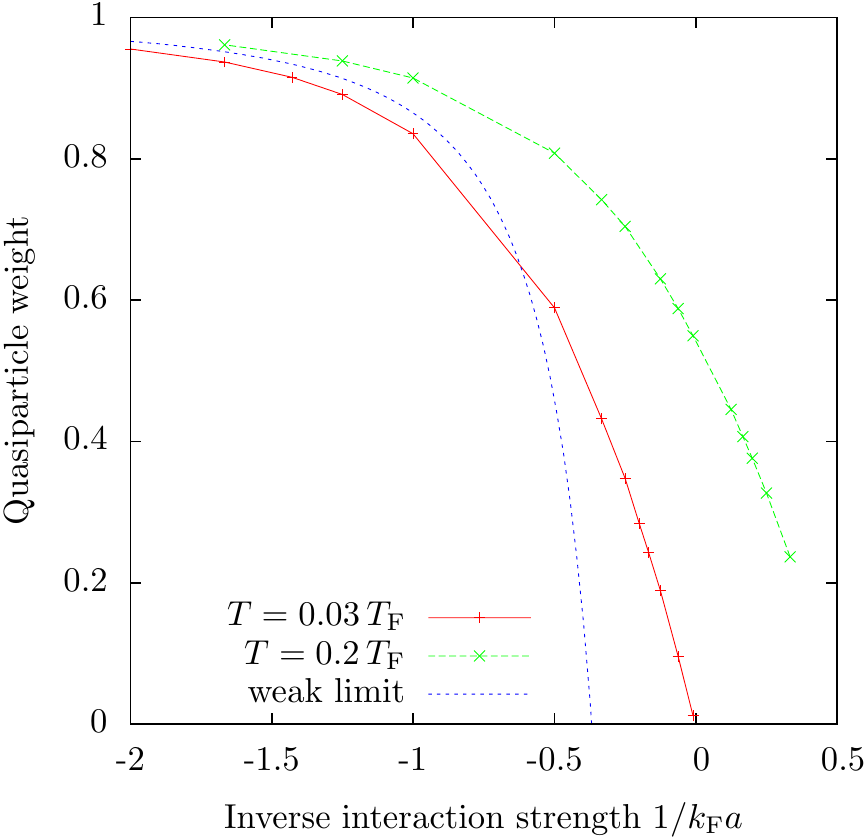}
\caption{Quasiparticle weight $Z$ as a function of interaction strength calculated from the momentum distributions for temperatures $T=0.03\,T_\mathrm{F}$ and $T=0.2\,T_\mathrm{F}$. Shown is also analytical result for zero temperature, valid in the weakly interacting limit.}
\label{fig:qpweight}
\end{figure}

Momentum distributions are plotted in Fig.~\ref{fig:momdist} for various interaction strengths. 
The height of the momentum distribution step at the Fermi surface can be associated with the quasiparticle weight $Z$.
However, at finite temperatures, thermal excitations broaden the Fermi surface, and an alternative way for characterizing $Z$ is needed. 
We determine $Z$ by calculating the largest depletion and the largest increase in the momentum distribution compared to the Fermi-Dirac distribution $n_k$.
In practice this means calculating the maximum $\delta n_\mathrm{max}$ and the minimum $\delta n_\mathrm{min}$ of the occupation number correction $\delta n_k = n(k)-n_k$. 
The quasiparticle weight $Z$ is then $1-\delta n_\mathrm{max}+\delta n_\mathrm{min}$. 
For a noninteracting system $Z$ defined as above is equal to 1, regardless of the temperature.
At zero temperature and finite interaction, $Z$ is equal to the step in the momentum distribution at the Fermi surface, thus reproducing the expected behavior of a Fermi liquid.

Fig.~\ref{fig:qpweight} shows these calculated quasiparticle weights as a function of interaction strength. 
Also plotted is an analytical zero-temperature result valid for weak repulsive interactions~\cite{FetterAndWalecka,Sartor1980a}:
\begin{equation}
Z_\mathrm{weak} = 1 - \frac{4}{3\pi^2} \left(k_\mathrm{F}a \right)^2.
\end{equation}
Our model reproduces this analytical result exactly in the weakly interacting limit.
Our model predicts a larger quasiparticle weight at unitarity than observed in the experiment~\cite{Sagi2014a}.
However, the theory does describe the qualitative behavior correctly, especially that the quasiparticle weight vanishes slightly on the repulsive side of the crossover.
The quasiparticle weight, particularly in the strongly interacting regime, depends rather strongly on the temperature, so the discrepancy with the experiment could partially be due to difficulties in precisely determining the temperature, but also due to the different schemes of determining the quantity $Z$.

The momentum distribution also yields the correct $k \rightarrow \infty$ asymptote. For large $k$ we get
\begin{equation}
  n(k) \approx \int \frac{d{\bf p}d{\bf q}}{(2\pi)^6}\left|\Gamma(p+q,\varepsilon_p + \varepsilon_q)\right|^2 \frac{n_p n_q}{\left(2\epsilon_k\right)^2}.
\label{eq:momasym}
\end{equation}
This asymptotic behavior is clearly shown in the logarithmic plot in Fig.~\ref{fig:momdist}. 
The prefactor of the $k^{-4}$ tail is called the contact parameter $C$, and from Eq.~\eqref{eq:momasym} we get
\begin{equation}
 C = \frac{m^2}{\hbar^4}V \int \frac{d{\bf p}d{\bf q}}{(2\pi)^6} \left|\Gamma(p+q,\varepsilon_p + \varepsilon_q)\right|^2 n_p n_q,
\end{equation}
where $V$ is the volume. 
The same result was obtained in Ref.~\cite{Kinnunen2012a} using a different approach. 

\begin{figure}[t]
\centering
\includegraphics[width=0.9\columnwidth]{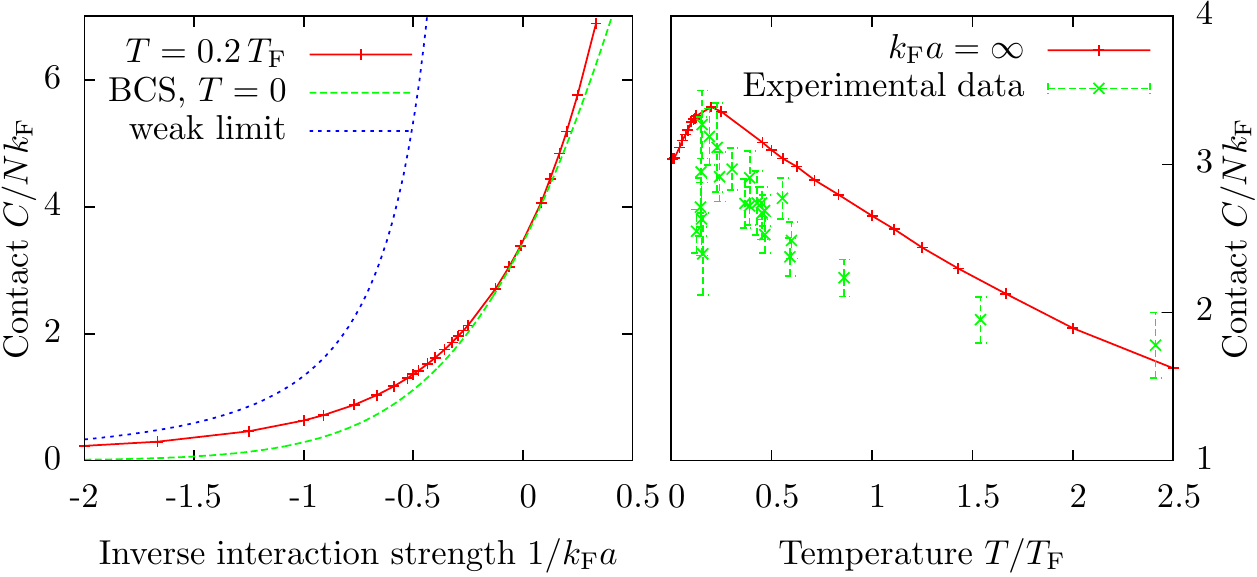}
\caption{Contact as a function of interaction strength for temperature $T=0.2\,T_\mathrm{F}$, and as a function of temperature at unitarity. Shown is also the analytical result $C/Nk_\mathrm{F} = \frac{4}{3} \left(k_\mathrm{F}a\right)^2$, which is valid for weak interactions. The numerical values reproduce the analytical result well at weaker interactions. Experimental data are from Ref.~\cite{Jin2012a}.}
\label{fig:contact}
\end{figure}

Fig.~\ref{fig:contact} shows the calculated contact as a function of interaction strength for $T=0.2\,T_\mathrm{F}$ and as a function of temperature at unitarity.
For weak interactions, the contact is given by the analytical result
\begin{equation}
  \frac{C_\mathrm{weak}}{Nk_\mathrm{F}} = \frac{4}{3} \left(k_\mathrm{F}a\right)^2.
\end{equation}
As is already well known, BCS theory is unable to reproduce this limit, but instead it predicts an exponentially decreasing contact as a function of interaction strength. 
The present theory reproduces the weak interaction result exactly. 

The temperature dependence of the contact shows a clear maximum close to the critical temperature for superfluidity, $T_\mathrm{c} \approx 0.2\,T_\mathrm{F}$, in qualitative agreement with predictions of an increase in the contact as a function of temperature for low temperatures \cite{Drut2011a, Yan2013a}. 
While the present model neglects superfluid properties, it produces well-behaved results for the contact parameter even in the low temperature regime.
For temperatures $T \gtrsim 0.2\,T_\mathrm{F}$ the contact decreases again. 
This is because the scattering T-matrix is strongly momentum dependent at unitarity, and the average relative momentum of the atoms increases with the temperature. 
The high-temperature limit reproduces the second order virial theorem result \cite{Hu2010b,Liu2013a}
\begin{equation}
 \frac{C_\mathrm{virial}}{Nk_\mathrm{F}} = 3\pi \left( \frac{T}{T_\mathrm{F}}\right)^2 z^2, 
\end{equation}
where $z = e^{-\mu/k_\mathrm{B}T}$.

{\bf Momentum-resolved radio-frequency spectroscopy.} 
The perturbative correction to the Green's function, $G_\mathrm{pert}$, allows also the study of momentum-resolved radio-frequency spectra. 
The momentum resolved spectrum, $S_k(\delta)$, consists of a bare response and the perturbative correction.
The former describes the response of the unperturbed propagator $G_\mathrm{BG}$:
\begin{equation}
  S_k^0(\delta) = \frac{2\eta_\mathrm{RF}}{\delta_k^2 + \eta_\mathrm{RF}^2},
\label{eq:barespec}
\end{equation}
where $\eta_\mathrm{RF}$ is the linewidth of the radio-frequency field, and
$\delta_k = \epsilon_k - \delta - \varepsilon_k$. Notice, that this response already contains the momentum-dependent Hartree-type energy shift through $\varepsilon_k = \epsilon_k + \Sigma_\mathrm{BG}(k)$.

The perturbative correction to the response function is derived in Section~\ref{sec:methods}, but it can be most easily described using schematic diagrams, shown in Fig.~\ref{fig:diagrams}. 
In the diagram A, before the absorption of the photon, the particles in the scattered states are simple virtual excitations with the energy of the scattered state being equal to the initial energy of the $+$ and $-$-atoms.
In order for the rf-photon to be absorbed, and the $k$ momentum atom being transferred to the excited $|e\rangle$-spin state, the photon will need to supply the required energy to make the virtual state real.
Hence the process is on-resonance at frequency $\delta = \epsilon_k + \varepsilon_p - \varepsilon_+ - \varepsilon_-$, corresponding to the increase in the kinetic energies due to the scattering, $\Delta E = \varepsilon_p + \varepsilon_k - \varepsilon_+ - \varepsilon_-$, and the energy change due to the absorption of the radio-frequency photon $\delta_k = \epsilon_k - \delta - \varepsilon_k$.

If there is a possibility of finding atoms at high momentum states, as described by the diagram A, the probability of finding atoms in low momentum states must decrease. 
This is indeed the effect of the diagram B in Fig.~\ref{fig:diagrams}.
The diagram provides a spectral response which has the same overall functional form as the bare response, $S_k^0(\delta)$, and since it describes a vacancy, it has the opposite sign.

The process described in the diagram C involves dynamics generated by the creation of the hole excitation~\cite{Leskinen2010a}, and it does not influence ground state properties, such as the momentum distribution.

\begin{figure}
\centering
\includegraphics[width=0.9\columnwidth]{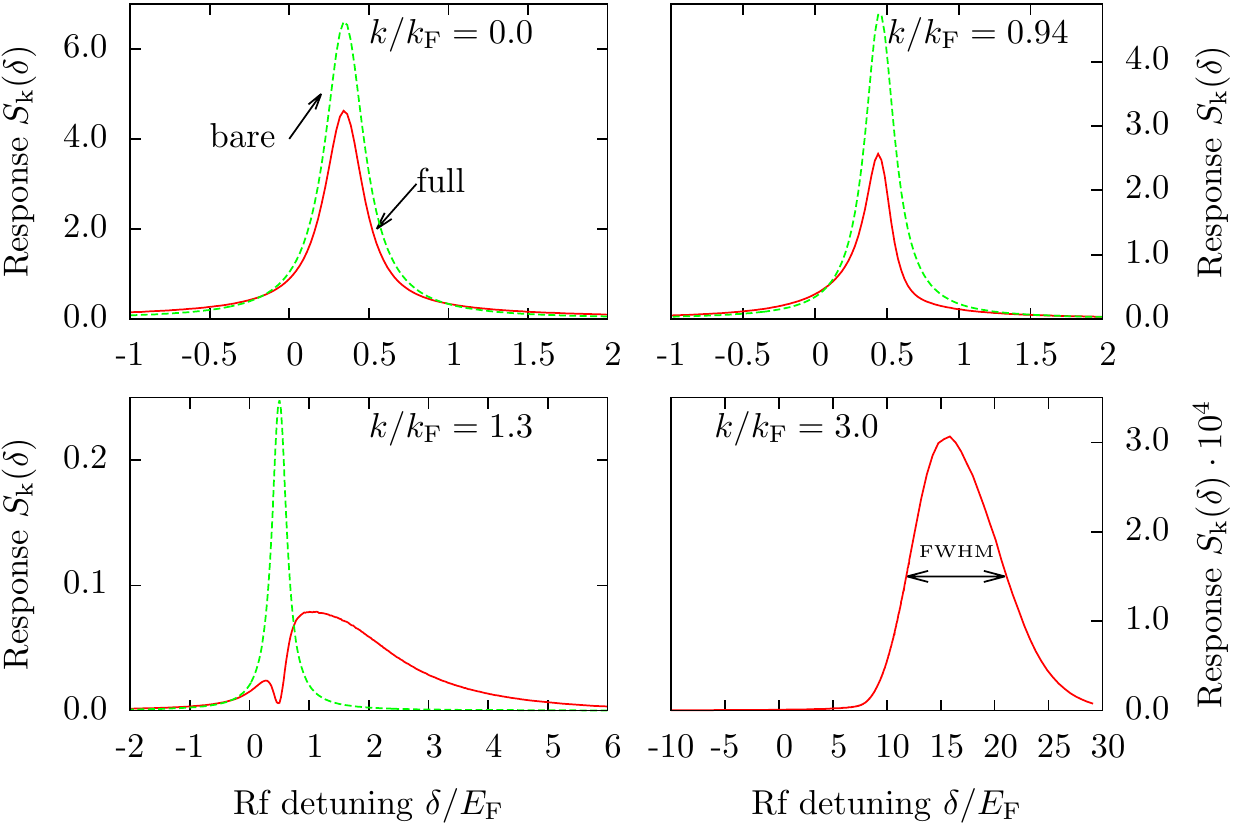}
\caption{Momentum-resolved radio-frequency spectral functions $S_k(\delta)$ for atoms with different momenta $k$. Top figures: for low momenta $k < k_\mathrm{F}$ the full response exhibits a significantly weaker response than the bare response, which involves only the unperturbed Brueckner-Goldstone contribution (Hartree shift). The corrections do not cause any additional shift but lower the peak and create a very broad incoherent background response (stronger tails). Bottom figures: for larger momenta $k > k_\mathrm{F}$, the bare response vanishes rapidly as only thermal quasiparticle excitations contribute to the unperturbed response. However, the full response acquires a very broad asymmetric peak, corresponding to scattered atoms. The calculated width (FWHM) of the spectral peak shown for $k=3\,k_\mathrm{F}$ data is $9.2\,E_\mathrm{F}$. Notice the additional factor $10^4$ included in the $k=3\,k_\mathrm{F}$ plot. Here, and in all the response data, the interaction strength is $k_\mathrm{F}a = -4$ and temperature $T = 0.2\,T_\mathrm{F}$.}
\label{fig:spectrals}
\end{figure}

Fig.~\ref{fig:spectrals} shows the momentum-resolved spectra calculated for different momenta. 
For hole excitations, $k < k_\mathrm{F}$, the second-order correction to the spectrum lowers the Lorentzian bare response peak significantly and creates a broad background response. 
Due to the weakness of the background response, the full width at half maximum (FWHM) of the full response is unaffected by the corrections.

The broad background response originates from the diagram C in Fig.~\ref{fig:diagrams}.
The scattering of the hole can decrease or increase the energy of the atoms, thus providing a resonant total process at energies significantly away from the single-particle resonance.
Also the diagram B in Fig.~\ref{fig:diagrams} affects low momentum atoms. 
However, since the contribution has exactly the same lineshape as the bare response, it can only lower the spectral peak by the amount corresponding to the quantum depletion of the momentum distribution.

Fig.~\ref{fig:spectrals} shows also the momentum-resolved spectrum for an atom with momentum $k=1.3\,k_\mathrm{F}$. At this momentum, there are still some thermal quasiparticle excitations, and consequently the bare response still appears.
In the full response, this quasiparticle peak is broadened and lowered, but in addition there appears a very broad and highly asymmetric feature.
For even higher momentum, $k=3\,k_\mathrm{F}$, the bare response is completely absent, since the momentum state is not populated in the unperturbed state (the Fermi-Dirac occupation probability is vanishingly low). 
However, the full response still exhibits a very broad spectral peak. 
The response, and the broad feature in the $k=1.3\,k_\mathrm{F}$ data, comes from the diagram A in Fig.~\ref{fig:diagrams}.
The radio-frequency field will need to supply the required energy to make the virtual excitation real.
However, since the transferred atom with momentum ${\bf k}$ may have reached the scattering state through interaction with any of the atoms in the $|\mydown\rangle$-Fermi sea, the virtual state has a very broad energy spectrum.

\begin{figure}
\centering
\includegraphics[width=0.9\columnwidth]{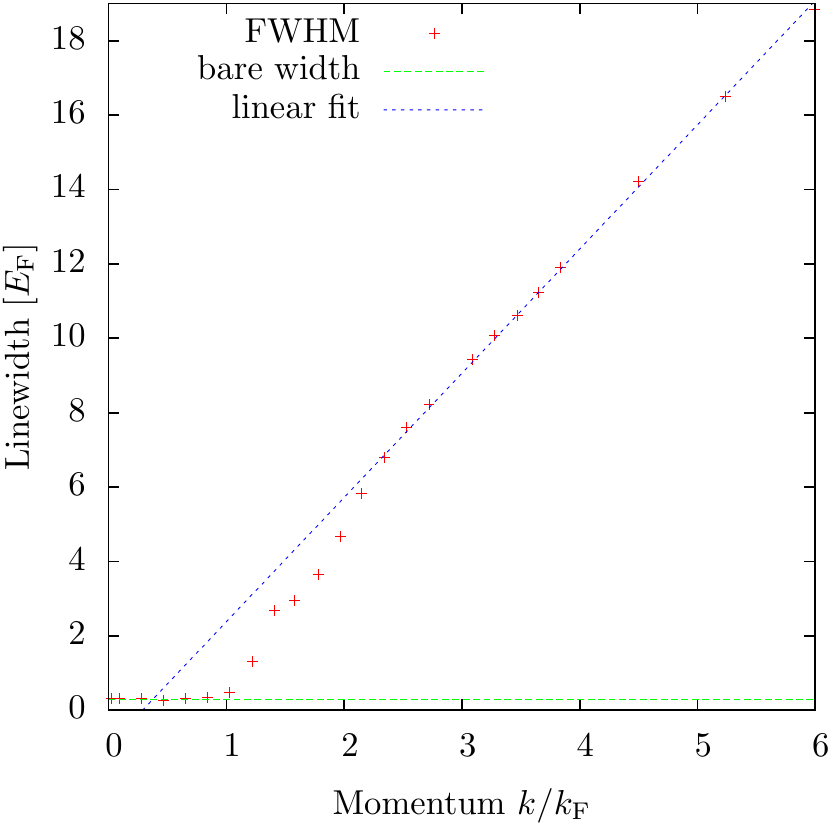}
\caption{The widths (full width at half-maximum) of the momentum-resolved spectra as a function of momentum. At low momenta, the width is dominated by the constant linewidth of the bare response (here $0.3\,E_\mathrm{F}$), but at high momenta the width increases linearly with momentum. The linear fit has slope $3.34\,E_\mathrm{F}/k_\mathrm{F}$ and is calculated for data $k/k_\mathrm{F} \in [2,6]$.}
\label{fig:rfwidth}
\end{figure}

Fig.~\ref{fig:rfwidth} shows the full-width at half-maximum (FWHM) of the spectral peaks as a function of momentum $k$. 
While FWHM is too rough a measure for revealing the effect of the hole dynamics at low momenta, which produces the broad incoherent background response seen in low momentum data in Fig.~\ref{fig:spectrals}, it does show a linear increase of the spectral width at high momenta. 
The increase is in drastic contrast with the width of the spectral peak predicted by BCS theory, which yields a spectral width limited only by the linewidth of the radio-frequency field.

If one interprets the results as a signature of bound pairs, the width of the peak can be understood as a measure of the imprecision in the center-of-mass momenta of pairs. 
Indeed, consider a bound pair of center-of-mass momentum $q$.
It can be described by the pair creation operator
\begin{equation}
   \hat \psi^\dagger_{\bf q} = \sum_{\bf k} v_{\bf k} \hat c_{{\bf k},\uparrow}^\dagger \hat c_{{\bf -k+q},\downarrow}^\dagger.
\end{equation}
Performing momentum-resolved spectroscopy for a pair created by such operator yields the momentum-resolved spectral function
\begin{align}
 S_{k,q}^\mathrm{pair}(\delta) & \sim \delta (-\delta + \epsilon_{\bf -k+q}+\epsilon_k+\Delta) \nonumber \\
 & = \delta (-\delta + 2\frac{\hbar^2 k^2}{2m} - 2\frac{\hbar^2}{2m} {\bf k}\cdot {\bf q} + \frac{\hbar^2 q^2}{2m} + \Delta),
\end{align}
where $\Delta$ is the pair binding energy (the initial energy of the pair).
The spectral function is thus a narrow peak at frequency $\delta = 2\frac{\hbar^2 k^2}{2m} - 2\frac{\hbar^2}{2m} {\bf k}\cdot {\bf q} + \frac{\hbar^2 q^2}{2m} + \Delta$. 

If there is spread in the center-of-mass momenta of the pairs, the spectral peak becomes broader. 
For example, if pairs have center-of-mass momenta in the interval $q \in [-q_c,q_c]$, the width of the spectral function is
\begin{align}
 &\Delta S = \left[2\frac{\hbar^2 k^2}{2m} + 2\frac{\hbar^2}{2m} kq_c + \frac{\hbar^2 q_c^2}{2m} + \Delta \right] \nonumber \\
 & - \left[2\frac{\hbar^2 k^2}{2m} - 2\frac{\hbar^2}{2m} kq_c + \frac{\hbar^2 q_c^2}{2m} + \Delta \right] = 4 \frac{\hbar^2}{2m} kq_c.
\end{align}
The width of the observed spectrum is thus a function that increases linearly with the momentum $k$ and the slope is given by the center-of-mass momentum spread of the pairs.

Considering the widths in Fig.~\ref{fig:rfwidth}, the corresponding pair center-of-mass momenta would be of the order of the Fermi momentum: a linear fit to data in the interval $k/k_\mathrm{F} \in [2,6]$ gives a slope of $3.34$, translating into a center-of-mass momentum width $q_c \approx 0.84\,k_\mathrm{F}$. 
The data in Fig.~\ref{fig:rfwidth} was calculated for interaction strength $k_\mathrm{F}a=-4$ and temperature $T=0.2\,T_\mathrm{F}$. 
This slope can be compared with the fitted pair temperature $T_\mathrm{P}$ observed in the experiment~\cite{Sagi2014a}, which is approximately $T_\mathrm{P} \approx 0.8\,T_\mathrm{F}$ throughout the BCS-BEC crossover. 
The observed pair temperature appears unrelated to the actual gas temperature $T \approx 0.25\,T_\mathrm{F}$.

Since the present theory can quantitatively describe the observed pair temperature without including bound pairs in the theoretical description~\cite{Perali2011a}, one can ask how actual condensed pairs would show up in the spectrum.
Considering the condensation of zero-momentum Cooper pairs in the superfluid phase, we expect to observe a narrow spectral feature in the high momentum momentum-resolved spectrum $S_k(\delta)$ when the temperature is reduced below the critical temperature.

\begin{figure}[!htb]
\centering
\includegraphics[width=0.9\columnwidth]{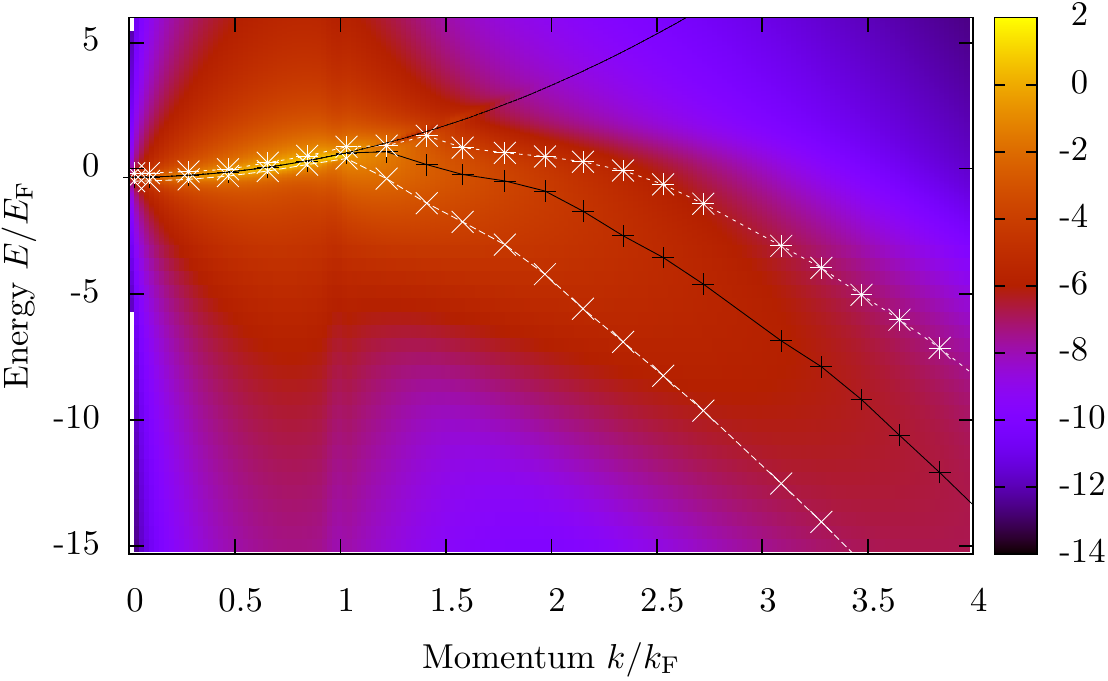}
\caption{Momentum-resolved radio-frequency spectra $k^2S_k(\delta)$. The $k^2$-prefactor provides the effect of the density of states, to provide easier comparison with experimental data. The colour bar shows the magnitude of the response in a logarithmic scale. 
Shown are also the position of the spectral maximum (black crosses) and the frequencies at which the response is half of the maximum value (white symbols) -- the full width at half-maximum is then the energy separation of the two half-maximum energies, used in Fig.~\ref{fig:rfwidth}.
The solid black line is a quadratic fit to the quasiparticle branch $E = E_0 + \frac{\hbar^2 k^2}{2m^*}$, with $E_0 = -0.33\,E_\mathrm{F}$ and $m^* = 1.1\,m$.}
\label{fig:momres_t20}
\end{figure}

Fig.~\ref{fig:momres_t20} shows the full momentum resolved rf-spectrum.
At low momenta $k < k_\mathrm{F}$ the response has quite a narrow linewidth, but at higher momenta a broad back-bending branch appears~\cite{Schneider2010a}.
Thermal excitations show up as a narrow quasiparticle branch extending beyond momenta $k > k_\mathrm{F}$. 
Notice, that many pseudogap theories~\cite{Chen2009b, Magierski2009a, Tsuchiya2010a, Palestini2012a}, 
exhibit an additional spectral branch at low momenta $k < k_\mathrm{F}$ and at positive energies $E >0$.
This branch is a remnant of the thermally excited quasiparticle branch present already in BCS theory but also in the fully self-consistent field theory in the superfluid phase~\cite{Haussmann2009a}.
It is noteworthy that the branch is missing in the present theory, but it is also missing from the experimental spectra~\cite{Sagi2014a}.

The momentum resolved spectrum is more sensitive than the momentum distribution to the perturbative corrections.
Indeed, the momentum distribution is well behaved across the BCS-BEC crossover for sufficiently high temperatures.
In contrast, the momentum resolved spectrum has artifacts near the Fermi surface, such as areas where the response becomes negative. 
This happens when the perturbative correction becomes larger than the unperturbed value, signaling a breakdown of the perturbative approach.
The reason these artifacts do not appear in momentum distribution is that the perturbative correction terms partially cancel each other.
However, since the different processes (or diagrams) in the perturbative correction are resonant at different energies, the partial cancellation does not happen when the energies are resolved, such as in the response function.
We are thus limited to weaker interactions in the momentum-resolved spectroscopy.

\begin{figure}[!htb]
\centering
\includegraphics[width=0.9\columnwidth]{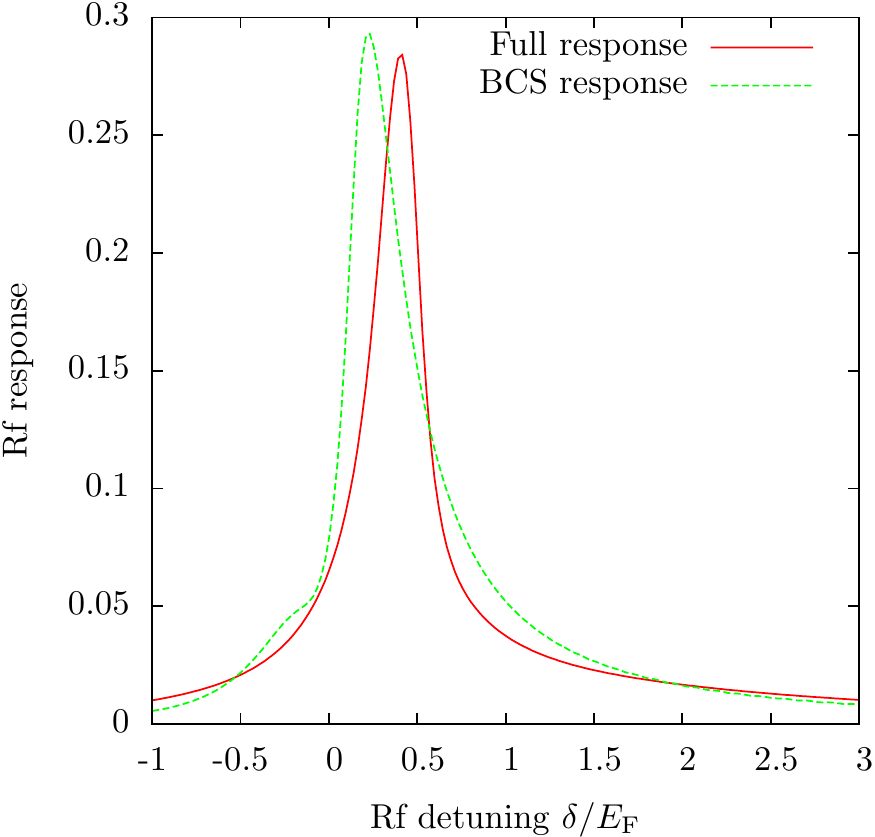}
\caption{Radio-frequency spectrum obtained by integrating the momentum resolved spectra over momentum $k$. Shown is also the spectrum obtained from BCS theory.
The main differences between the two spectra are the slightly more asymmetrical lineshape and a broad bump at negative detuning $\delta \approx -0.3\,E_\mathrm{F}$ in the BCS response. The latter feature comes from thermal quasiparticle excitations present in the BCS theory~\cite{Kinnunen2004a}. However, the uniform rf-response is expected to have only a single peak~\cite{Massignan2008a}.}
\label{fig:rf}
\end{figure}

Fig.~\ref{fig:rf} shows the calculated (non-momentum-resolved) radio-frequency spectrum and the corresponding result from BCS theory, obtained from the momentum-resolved spectrum by integrating over the momentum $k$. 
Even though the two theories yield qualitatively different momentum-resolved spectra, the two agree surprisingly well in the integrated response.
The interpretations of the two spectra in Fig.~\ref{fig:rf} are, however, quite different. 
While the energy shift in the BCS spectrum is due to pair binding energy $\Delta$, the present theory explains it as a simple Hartree-type energy shift.
It thus appears that the Hartree energy shift turns into pair binding energy~\cite{Chin2004a,Kinnunen2004b} when the transition from the normal phase to the superfluid phase occurs.
The Hartree shift is also the dominant energy shift in the weakly attractive regime, even at zero temperature~\cite{Haussmann2009a}.

The radio-frequency spectrum in Fig.~\ref{fig:rf} is in good qualitative agreement with experimental spectra for uniform systems~\cite{Shin2007a,Jin2012a}.
It will be very interesting to see how the present theory works with spin-imbalanced systems, and, particularly, whether the model can produce a double peak structure as observed in Ref.~\cite{Schirotzek2008a}. 
However, this goes beyond the scope of the present investigation.

\section{Discussion}

In conclusion, we have perturbatively extended the Brueckner-Goldstone (BG) theory and applied it to a strongly interacting Fermi gas in the BCS-BEC crossover.
The theory provides direct access to momentum distributions and momentum-resolved radio-frequency spectra.
The momentum distributions are consistent with exact asymptotic results from the Tan relations, giving a high-momentum tail with an algebraic $1/k^4$-decay.
Moreover, the strength of the algebraic decay is in good agreement with experimentally determined values.
We also find good agreement between the radio-frequency spectra predicted from the extended BG theory and experimental spectra.
Furthermore, we predict the breakdown of Fermi liquid behavior at finite repulsion, also in agreement with the JILA experiment~\cite{Sagi2014a}.

The model used here neglected possible bound pairs in order to help formulate a theory in line with Fermi liquid theory.
However, it is important to notice that the model does not exclude pair correlations.
Indeed, the perturbative correction to Brueckner-Goldstone theory can be understood as introducing pair correlations, that were lost in the on-shell approximation of the self-energy.
Many properties of the model studied here, particularly the back-bending part of the high momentum momentum resolved spectrum, can be understood as a signature of these correlations~\cite{Schneider2010a}. 
But pair correlations are unavoidable in interacting systems, and have very little to do with presence of bound pairs.

The contact interaction potential used in this work describes Fermi gases with so called broad Feshbach resonances well. 
However, the one used in Ref.~\cite{Sagi2014a} involved a narrow Feshbach resonance, for which one would ideally like to include also effective range correction in the theory~\cite{Parish2005a,Jensen2006a}.
While we do not expect this oversight to change results qualitatively, one would expect that the inclusion of the effective range correction would be needed for a quantitatively accurate description of the system.

The present work points out several quantities that could be studied in the experiments, such as the broad incoherent background response in the 
momentum-resolved rf-spectrum at low momenta, the asymmetry of the spectral linewidth, and the linear scaling of the width of the high momentum response peak.
Furthermore, we expect the transition to the superfluid state, with condensed pairs, to be reflected as a narrowing or at least as an emergence of some narrow feature in the high momentum radio-frequency response.

\section{Methods}
\label{sec:methods}
In this section we will first present the self-consistent Brueckner-Goldstone theory and then explain how it is extended using perturbation theory.
The result is a theory that can describe Hartree-type energy shifts (although momentum dependent) and predicts qualitatively correct behavior for the asymptotic momentum distribution.
In the weakly interacting limit, the theory reproduces well-known analytical results, but the model is well-behaved also across the BCS-BEC crossover.

For a spin-balanced system and equal masses, the Green's functions and self-energies for both $|\myup\rangle$- and $|\mydown\rangle$-spin states are identical. 
Hence for simplicity, we will consider only the Green's function and self-energy of the $|\myup\rangle$ spin state.
However, one should keep in mind that the Green's functions considered below describe only the two interacting spin components, $|\myup\rangle$, $|\mydown\rangle$, whereas the Green's function for the excited $|e\rangle$-spin state is the one expressed in Eq.~\eqref{eq:nonintG}.

The many-body (dressed) Green's function can be calculated from the Dyson equation (using the four-vector notation $K = ({\bf k},\omega)$)
\begin{equation}
  G_\uparrow(K)^{-1} = G_T(K)^{-1} - \Sigma_\uparrow(K).
\end{equation}The non-interacting finite temperature Green's function at temperature $T$ is given by $G_T(K) = \frac{n_k}{\omega - \epsilon_k - i\eta} + \frac{1-n_k}{\omega - \epsilon_k + i\eta}$, where $n_k = f(\epsilon_k) = 1/(1+e^{\beta (\epsilon_k-\mu)})$ is the Fermi-Dirac distribution, $\mu$ is the chemical potential and $\beta = 1/k_BT$. 
This non-interacting Green's function $G_T(K)$ describes both hole (first term) and particle (second term) excitations in the thermal Fermi sea, but neglects any interaction effects. 
These effects enter the dressed Green's function $G_\uparrow(K)$ through the self-energy $\Sigma_\uparrow(K)$, which 
in the ladder approximation is
\begin{equation}
\Sigma_\uparrow (K) = \int \frac{dP}{i(2\pi)^4} \Gamma(K+P) G_\downarrow(P).
\label{eq:selfenergy}
\end{equation}
Here $\Gamma(K)$ is the many-body scattering T-matrix
\begin{equation}
  \Gamma(K) = \frac{\Gamma_0(K)}{1+\Gamma_0(K) \left(\chi(K)-\chi_0(K)\right)},
\end{equation}
where the pair susceptibility $\chi(K) = \int \frac{dQ}{i(2\pi)^4} G_\uparrow(K+Q/2) G_\downarrow(K-Q/2)$ and the two-body scattering
T-matrix $\Gamma_0(K) = \Gamma_0({\bf k},\omega) = V_0 (1+a\sqrt{\omega-\epsilon_k})^{-1}$ \cite{Morgan2002a}. 
Expressing the many-body scattering T-matrix using the two-body scattering T-matrix involves double counting certain scattering terms. 
In order to remove these artifacts, one needs to remove the vacuum pair susceptibility $\chi_0(K) = \int \frac{dQ}{i(2\pi)^4} G_0(K+Q/2) G_0(K-Q/2)$  
from the pair susceptibility $\chi(K)$.

In the Brueckner-Goldstone theory, the frequency dependence of the self-energy is neglected and the value of the self-energy is evaluated on-shell. 
That is, the self-energy entering the Brueckner-Goldstone Green's function $G_\mathrm{BG}(K)^{-1} = G_T(K)^{-1} - \Sigma_\mathrm{BG}(k)$ is solved iteratively as
\begin{equation}
  \Sigma_\mathrm{BG} (k) = \Sigma_\uparrow(k,\epsilon_k + \Sigma_\mathrm{BG}(k)).
\end{equation}
The theory is fully self-consistent, in the sense that the Brueckner-Goldstone Green's function $G_\mathrm{BG}(K)$, obtained from the Dyson equation, is used in the pair susceptibility $\chi(K)$ and the self-energy $\Sigma_\uparrow(K)$.

Due to the simplicity of the self-energy, the Brueckner-Goldstone Green's function also has a very simple form at finite temperatures
\begin{align}
  &G_\mathrm{BG}({\bf k},\omega) = \frac{n_k}{\omega-\epsilon_k - \Sigma_\mathrm{BG}(k) - i\eta} \nonumber \\
  &+ \frac{1-n_k}{\omega-\epsilon_k-\Sigma_\mathrm{BG}(k)+i\eta}.
\label{eq:bggreen}
\end{align}
Notice that the Brueckner-Goldstone self-energy will not affect the momentum distribution and thus it is sufficient to solve the distribution $n_k$ for the noninteracting system when fixing the number of atoms in the system.
This does not mean that the Brueckner-Goldstone self-energy does not affect the chemical potential $\mu$: expressing the Fermi-Dirac distribution in terms of the interacting single-particle energies $\epsilon_k + \Sigma_\mathrm{BG}(k)$ shows that the true chemical potential of the interacting system is $\mu + \Sigma_\mathrm{BG}(k_\mathrm{F})$.

The real part of the Brueckner-Goldstone self-energy $\mathrm{Re} \, \Sigma_\mathrm{BG}(k)$ can be interpreted as the Hartree energy shift since in the weakly interacting 3d limit it yields the standard result $\frac{4\pi\hbar^2}{m} n_\sigma a$, where $n_\sigma$ is the atom density in spin state $|\sigma\rangle$. However, the energy shift depends on the momentum 
because of the momentum dependence of the scattering T-matrix. The imaginary part $\mathrm{Im} \, \Sigma_\mathrm{BG}(k)$ has correct Fermi liquid features so that for $k > k_\mathrm{F}$ the imaginary part is negative, corresponding to particle excitations, and for $k < k_\mathrm{F}$ the imaginary part is positive as required for hole excitations. 
At the Fermi surface the imaginary part vanishes, which is a signature that the Brueckner-Goldstone theory provides well-defined quasiparticles.
In principle, the auxiliary convergence parameter $\eta$ is not needed, because the imaginary part of the self-energy itself could provide a necessary convergence factor.

While the Brueckner-Goldstone theory is self-consistent, it is unable to describe pairs.
The pair formation is caused by the presence of poles in the scattering T-matrix, and it appears in the self-energy landscape $\Sigma(k,\omega)$ as a peak along the $\omega \approx \Delta -\frac{\hbar^2 k^2}{2m}$-branch, where $\Delta$ is the pair binding energy.
This branch is missed by the Brueckner-Goldstone self-energy $\Sigma_\mathrm{BG}$. 
Since the pair formation cannot therefore be self-consistently described, we make a further approximation and neglect poles in the many-body scattering T-matrix. 
In practice, this is performed by replacing the many-body scattering T-matrix by the on-shell T-matrix.
The Brueckner-Goldstone self-energy now acquires a particularly simple form:
\begin{equation}
  \Sigma_\mathrm{BG}^\mathrm{os}({\bf k}) = \int \frac{d{\bf p}}{\left(2\pi\right)^3} n_p \Gamma({\bf k + p},\varepsilon_k + \varepsilon_p),
\label{eq:brsigma}
\end{equation}
where $\varepsilon_k = \epsilon_k + \Sigma_\mathrm{BG}^\mathrm{os}(k)$.

Neglecting the poles of the scattering T-matrix $\Gamma$, however, breaks the analytical structure of the equation and results in unphysical functional dependence of the imaginary part of the self-energy. 
In particular, the imaginary part of the self-energy no longer changes sign at the Fermi surface.
In Ref.~\cite{Kinnunen2012a}, one of us used a model in which the correct sign was imposed on the imaginary part by hand. 
However, this approach still has the problem that the value of the imaginary part in the vicinity of the Fermi surface
is anomalously large (indeed, it had a maximum near the Fermi surface). 
Here we use an even simpler approach: we neglect the imaginary part obtained from the Brueckner-Goldstone self-energy altogether and instead use a fixed imaginary part. 
While this means that quasiparticle excitations at the Fermi surface still have a finite lifetime, it does not have any qualitative effect in the results shown below. 
We have checked this by trying alternative schemes in which the imaginary part crosses zero continuously at the Fermi surface, that is, $\eta_k \sim (k-k_\mathrm{F})^\alpha$ with $\alpha=1$ or $\alpha=3$.
However, while the value of $\eta$ in the vicinity of the Fermi surface is not important, the overall value of $\eta$ does affect the results to some extent.
Throughout this work, we use the value $\eta=0.05\,E_\mathrm{F}$, corresponding roughly to the imaginary part of the Brueckner-Goldstone self-energy at zero momentum for interaction strength $k_\mathrm{F}a = -2$, i.e. $\eta = \mathrm{Im}\, \Sigma_\mathrm{BG}(k=0)$.
While different choices of parameter $\eta$ do not result in any qualitative changes, the actual numerical values of the contact, effective mass and quasiparticle ratio change by at most $10\,\%$ when $\eta$ is decreased by factor $0.5$ or increased by factor $1.5$. 
As expected, largest effect is found at unitarity and low temperatures, where the T-matrix is most strongly peaked.
At weaker interactions and/or higher temperatures the results are less sensitive to the value of $\eta$.

A minimalistic way to correct the problem with the sign of the imaginary part, would be to include in the self-energy all terms of second order in the on-shell T-matrix $\Gamma_\mathrm{on}$.
This would make the theory formally similar to the Galitskii's theory~\cite{FetterAndWalecka}, but with the vacuum scattering amplitude $\frac{4\pi\hbar^2 a}{m}$ replaced by $\Gamma_\mathrm{on}$.
However, this would increase the numerical complexity of the present theory, and is thus not done here.

The self-consistent Brueckner-Goldstone self-energy $\Sigma_\mathrm{BG}(k)$, and the associated Green's function $G_\mathrm{BG}(K)$, provide a good basis for a perturbative expansion. 
Indeed, as shown analytically in Ref.~\cite{Mahaux1992a}, the expansion done in Eq.~\eqref{eq:pertG} satisfies the Migdal-Luttinger theorem \cite{Migdal1967a}, yielding a step in the zero-temperature momentum distribution at the Fermi surface and even satisfying number conservation.
Furthermore, the expansion allows calculating values of many physical observables, such as the momentum-resolved radio-frequency spectrum.

The spectrum is defined in Eq.~\eqref{eq:spectrum}. Using the perturbed Green's function defined in Eq.~\eqref{eq:pertG} we get
\begin{eqnarray}
 \nonumber S_k(\delta) &=& S_k^0(\delta) + \int \frac{d\omega}{2\pi i} G_e(k,\omega+\delta) \\
&\times& G_\mathrm{BG}(k,\omega)^2 \left[ \Sigma(k,\omega)-\Sigma_\mathrm{BG}(k)\right],
\label{app:eq:sk}
\end{eqnarray}
where $S_k^0(\delta)$ is the bare response in Eq.~\eqref{eq:barespec} and
the self-energy is defined in Eq.~\eqref{eq:selfenergy}.
The self-energy contains the many-body scattering T-matrix $\Gamma = \Gamma({\bf k+p},\omega+\Omega)$, which can be expressed in terms of the on-shell T-matrix $\Gamma_\mathrm{os}$. 
In the on-shell T-matrix, the frequencies $\omega$ and $\Omega$ are replaced by the energies $\varepsilon_k$ and $\varepsilon_p$ of the incoming (scattering) particles.
The many-body scattering T-matrix can now be written as
\begin{eqnarray}
    \Gamma &=& \frac{\Gamma_\mathrm{os}}{1-\Gamma_\mathrm{os} \left[ \chi({\bf k+p},\omega+\Omega) - \chi({\bf k+p},\varepsilon_k+\varepsilon_p)\right]}  \\
&\approx& \Gamma_\mathrm{os} + |\Gamma_\mathrm{os}|^2 \left[ \chi({\bf k+p},\omega+\Omega) - \chi({\bf k+p},\varepsilon_k+\varepsilon_p) \right].\nonumber
\end{eqnarray}
The response Eq.~\eqref{app:eq:sk} becomes now
\begin{widetext}
\begin{eqnarray}
  S_k(\delta) = S_k^0(\delta) &-& 2\mathrm{Im}\,\int \frac{d{\bf p}d\omega d\Omega}{(2\pi)^5} |\Gamma_\mathrm{os}|^2 G_e(k,\omega+\delta) G_\mathrm{BG}(k,\omega)^2 \left[ \chi({\bf k+p},\omega+\Omega) - \chi({\bf k+p},\varepsilon_k+\Omega) \right] G(p,\Omega) \nonumber \\
= S_k^0(\delta) &-& \int \frac{d{\bf p}d{\bf q}}{(2\pi)^6} |\Gamma_\mathrm{os}|^2 \frac{n_k n_p(1-n_+)(1-n_-)}{(\varepsilon_k+\varepsilon_p-\varepsilon_+-\varepsilon_-)^2}\frac{2\eta}{(\epsilon_k-\delta-\varepsilon_k)^2 + \eta^2} \\
&+& \int \frac{d{\bf p}d{\bf q}}{(2\pi)^6} |\Gamma_\mathrm{os}|^2 \frac{(1-n_k)(1-n_p)n_+n_-}{(\varepsilon_++\varepsilon_--\varepsilon_p-\varepsilon_k)^2} \frac{2\eta}{(\epsilon_k-\delta+\varepsilon_p-\varepsilon_+-\varepsilon_-)^2 + \eta^2}\nonumber \\
&-& \int \frac{d{\bf p}d{\bf q}}{(2\pi)^6} |\Gamma_\mathrm{os}|^2 \frac{n_k(1-n_p)n_+n_-}{\varepsilon_k+\varepsilon_p-\varepsilon_+-\varepsilon_-} \mathrm{Im}\,\frac{2}{(\epsilon_k-\delta+\varepsilon_p-\varepsilon_+-\varepsilon_--3i\eta)(\epsilon_k-\delta-\varepsilon_k-i\eta)},\nonumber
\end{eqnarray}
\end{widetext}
where $\pm$ indices refer to momenta $({\bf k+p})/2\pm {\bf q}$.

From the spectrum, we can obtain also the momentum distribution by integrating over detuning $\delta$. Simple algrebra leads into Eq.~\eqref{eq:momcorr}.
The same equation was obtained also in Ref.~\cite{Mahaux1992a} but there the many-body scattering T-matrix $\Gamma$ was the Brueckner's reaction matrix and the single-particle energies neglected the self-energy shift. The two correction terms to the momentum distribution cancel each other when integrated over the momentum $k$, satisfying thus the number conservation. 
Hence, also the 0-sum rule 
\begin{equation}
\int \frac{d{\bf k}}{(2\pi)^3} \int \frac{d\delta}{2\pi} \,S(\delta)_k = N
\end{equation}
is satisfied for the perturbed spectral function.

\begin{acknowledgments}
This work was supported by the Academy of Finland through its Centres of Excellence Programme (2012-2017) under Project No.\ 251748.
We are grateful to Y. Sagi for sharing experimental data.
\end{acknowledgments}

\bibliographystyle{apsrev4-1}
\bibliography{ref}

\end{document}